\documentclass[9pt,conference]{IEEEtran} 
\usepackage{cite}
\usepackage{amsmath,amssymb,amsfonts}
\usepackage{algorithmic}
\usepackage{graphicx}
\usepackage{textcomp}
\usepackage{url}
\usepackage{xcolor}
\def\BibTeX{{\rm B\kern-.05em{\sc i\kern-.025em b}\kern-.08em
    T\kern-.1667em\lower.7ex\hbox{E}\kern-.125emX}}

\usepackage{xspace}

{
\definecolor{ibcolor}{rgb}{0.4,0.6,0.2}
\definecolor{nkcolor}{rgb}{0.1,0.1,0.9}
\definecolor{todocolor}{rgb}{0.9,0.1,0.1}

\begin{document}

\title{Gridiron: A Technique for Augmenting Cloud Workloads\\
with Network Bandwidth Requirements}

\author{
{\rm Nodir Kodirov\textsuperscript{u},
Shane Bergsma\textsuperscript{h}},
Syed M. Iqbal\textsuperscript{u},
Alan J. Hu\textsuperscript{u},
Ivan Beschastnikh\textsuperscript{u},
Margo Seltzer\textsuperscript{u} \\
{\textsuperscript{u}University of British Columbia \hspace*{0.5cm}
\textsuperscript{h}Huawei Research, Canada}
}



\maketitle

\begin{abstract}
Cloud applications use more than just server resources,
they also require networking resources.
We propose a new technique to model network bandwidth demand of networked cloud
applications. Our technique, Gridiron, augments VM workload traces from Azure
cloud with network bandwidth requirements.
The key to the Gridiron technique is to derive inter-VM network bandwidth
requirements using Amdahl's second law.
As a case study, we use Gridiron to generate realistic traces with network
bandwidth demands for a distributed machine learning training application.
Workloads generated with Gridiron allow datacenter operators to estimate
the network bandwidth demands of cloud applications
and enable more realistic cloud resource scheduler evaluation.
\end{abstract}


\section{Introduction}
\label{sec:intro}

Datacenter resource allocation for networked cloud applications,
an area known as Virtual Datacenter (VDC) scheduling,
has been extensively studied over the past decade,
e.g.,~\cite{secondnet, oktopus, netsolver}.
A VDC includes a collection of VMs and the resource
requirements of each VM (e.g., CPU and RAM) as well as inter-VM
network bandwidth requirements~\cite{secondnet}.
However, existing works use synthetic workloads to evaluate VDC schedulers'
performance, because there is no publicly released production workload
with inter-VM bandwidth requirements.
We address this problem by proposing a new technique for generating
realistic VDC workloads. We call this the Gridiron technique.\footnote{The
term ``gridiron'' is borrowed from the urban planning literature. Just like
gridiron planning converts implicitly connected village houses into explicitly
connected network of urban residences, the Gridiron technique converts
implicitly connected tenant VMs into VDCs:
a network of explicitly connected VMs.}
In the Gridiron technique,
we use production traces from the Azure cloud as the
VM workload~\cite{rc} and augment its VMs with
network bandwidth requirements to generate a realistic VDC workload.

In generating a realistic VDC workload, we face the following question:
\textit{How much inter-VM network bandwidth should the VDC workload demand?}
To answer this question, we take inspiration from cloud computing history.
In the early days of cloud computing,
cloud providers faced the same question, but for the compute service.
For example, in 2006, public cloud pioneer Amazon had to decide
how much compute capacity to offer in a VM flavor in the EC2 product.
EC2 beta started with only one VM flavor that included 1 virtual CPU,
which offered an equivalent of Intel Xeon 1.7 GHz processor~\cite{ec2-beta}.
They also released several customer use-cases that might
benefit from this VM flavor,
such as a web-based back-office inventory application, and
three-tier web application (presentation tier, application tier, data tier),
which were popular applications at that time~\cite{ec2-beta}.
These use-cases are evidence that the decision for a VM's compute capacity
was driven by what EC2 cloud operators \textit{expected} tenants to run
and pay for in the cloud.

We expect that
VDC network bandwidth guarantees will evolve in the same way:
driven by what \textit{networked} applications providers expect
tenants to run and pay for in the cloud.
Note that this is different from today's networked cloud applications
for which free, best-effort networking suffices.
The networked cloud applications we believe ought to be run in VDCs
are the ones that will see significant performance benefit,
such as job completion time reduced by half,
from the bandwidth guarantee that the VDC abstraction offers.

An emerging cloud workload that was demonstrated to significantly benefit
from network bandwidth guarantees is distributed Machine Learning (ML)
training~\cite{p3, tictac, bytescheduler}.
ML training is typically deployed across multiple
\textit{communicating} VMs, or as a VDC,
where each VM trains an ML model on a subset of the data.
Existing literature show that these distributed ML training jobs complete
2--3$\times$ faster with consistent inter-VM network bandwidth guarantees, e.g.
P3~\cite{p3}, TicTac~\cite{tictac}, and ByteScheduler~\cite{bytescheduler}.

We will use distributed ML training as the sample VDC application
because it is emerging as one of the dominant cloud workloads~\cite{tiresias}.
Distributed training is an instance of parallel computing
with both \textit{sequential} and \textit{parallel} phases
\cite{parallel-computing}.
The sequential phase, which is called parameter aggregation,
is run by the parameter server, while the parallel phase, model training,
is split and run across multiple worker VMs.
The worker VMs synchronize their local state over the network.
Thus, we can apply Amdahl's second law~\cite{amdalh-2nd-law}
to estimate the network bandwidth requirements of this application.

Amdahl's second law, which is also called Amdahl's I/O law,
estimates the required network bandwidth
to have a \textit{balanced} system. A system is called balanced when it can
perform a compute task without memory or I/O bottlenecks
\cite{balanced-system, amdalh-2nd-law}.
The law states that for every one instruction per second of 
processing speed (Hz), a balanced computing system needs to provide
one bit per second of I/O rate (bps) and one byte of main memory capacity
\cite{balanced-system}.
For example, the initial VM flavor offered by EC2 beta with a 1.7 GHz processor
was a balanced system in terms of memory (1.75 GB of RAM),
but not in terms of I/O (250 Mbps)~\cite{ec2-beta}:\\
\centerline{$Amdahl~memory~number = \frac{memory~size~(GB)}{CPU~speed~(GHz)}
= 1.75/1.7 \approx 1$} \\
\centerline{$Amdahl~I/O~number = \frac{bandwidth~(Gbps)}{CPU~speed~(GHz)} = 0.25/1.7 \approx 0.15$}
Note that we use the network bandwidth, not the disk bandwidth, to derive the
Amdahl I/O number because our distributed ML training application
communicates parameter updates (from the worker VMs) to the parameter server VM
over the network.

We can use Amdahl's second law to answer our question:
\textit{VMs' network bandwidth should be proportional to their
compute capacity: 1 bit per second for every one CPU instruction in a
truly parallel VDC application}.
We call this the \textit{compute-proportional-bandwidth} approach for VDC
workload generation.
This approach is similar to how the Google Compute Engine (GCE)
scales its VM's \textit{egress} network bandwidth as a function of vCPUs
in the VM~\cite{gcp-network}.
Note that unlike VDCs in our model,
GCE's egress bandwidth is best-effort and is not between a pair of VMs.
Even though egress bandwidth is from an individual VM perspective,
the GCE example demonstrates the practicality of a
compute-proportional-bandwidth approach in a cloud environment.
Thus, we can generate a VDC workload from a VM workload
by using Amdahl's second law as our guiding principle.

An optimal Amdahl I/O number is deployment dependent, i.e.,
it depends on the characteristics of the cloud application and
the datacenter the application is running on.
For parallel applications, Amdahl's second law recommends a value of 1,
which is one datapoint in a spectrum.
Not all applications need a balanced system.
For example, Liang et al.\ demonstrate that an optimal Amdahl I/O number for
MPI-like (Message Passing Interface) applications ranges between 0.02--0.21,
while it ranges between 0.02--0.85 for Hadoop-like applications
\cite{balanced-system}.
Therefore, our generated VDC workload should be \textit{adaptable},
i.e., we should be able to adjust the network bandwidth demand
of the workload for the datacenter under test.
In section~\ref{sub:parameter}, we describe a parameterized
VDC workload generation methodology that allows adapting the VDC workload
to the datacenter under test.


\section{The Base Workload}
\label{sec:base}

We generate a VDC workload by augmenting the Azure cloud's production trace
with network bandwidth requirements. The Azure trace~\cite{dataset}
is released with the Resource Central paper~\cite{rc}.
It includes only VM CPU and RAM requirements.
We first we describe the characteristics of the Azure trace,
such as the number of VMs in the workload, and their CPU
and RAM footprints. Next, we describe how we use the deployment IDs,
which are included in the Azure trace, to group VMs into VDCs.

The Azure trace contains a list of 2,013,767 unique VMs in 35,941
unique \textit{deployments} over 30 days.
A deployment is defined as
``a set of VMs that the customer groups and manages together''~\cite{rc}.
For each VM, the dataset includes a VM creation and deletion time,
the deployment which it belongs to, the number of vCPUs and 
amount of RAM in the VM, and the vCPU utilizations in 5 minute intervals.
Here, all times are reported with 5 min granularity;
we call each such five minute interval a \textit{tick}.
The trace contains 8,640 ticks.

We preprocess the Azure trace to correct the VM timestamps.
First, we adjust VM create and delete timestamps to conform with
5-minute capturing interval.
There are 27 VMs with invalid create and delete timestamps.
We round invalid timestamps to the closest valid timestamp
as suggested by the Azure trace authors~\cite{timestamp}.
Second, we eliminate VMs that have identical creation and deletion tick,
which we call \textit{instant-VMs}.
There are 53,467 instant-VMs (less than 2\% of total),
which leaves 1,960,300 VMs in the preprocessed workload.
The Resource Central paper refers to these VMs as
``control plane latency test VMs''~\cite{rc};
these test VMs are used to continuously evaluate VM creation latency.

In summary, there are 2,013,767 VMs in 35,941 deployments before we preprocess
the Azure trace and 1,960,300 VMs in 35,870 deployments afterwards.
We call the Azure trace after preprocessing the \textit{base workload}.
The base workload's compute and memory footprint
varies around 10\% across all ticks:
it ranges between 321,043 cores and 346,755 cores, and
between 730,314 GB and 781,767 GB.
%
%

\section{From VMs to VDCs: The Gridiron Technique}
\label{sec:gridiron}

The first step towards generating a VDC workload is to decide how VMs
are grouped in the VDCs. Fortunately, the base workload already groups
VMs by ``deployment''.
When each deployment in the base workload is treated as the VDC,
there are 35,870 VDCs in the base workload.
The base workload dictates which VDC the VM belongs to and the tick
a VM joins the VDC (creation tick) and the tick a VM leaves the VDC
(deletion tick).

There are four additional steps in the Gridiron technique.
First, VDCs have inter-VM communication topologies, which the base workload lacks.
In section~\ref{sub:topologies}, we describe various topologies
VDCs can have and explain our rational for adopting an all-to-all topology.
Second, VDC sizes should be appropriate for the application(s) that these
VDCs encapsulate. In section~\ref{sub:vdc-size}, we explain the
significance of the peak VDC sizes, and show peak sizes in the base workload.
Third, virtual links (vlinks) in a VDC should have a bandwidth value.
In section~\ref{sub:parameter}, we use
the compute-proportional-bandwidth approach for assigning vlink bandwidths
and constructing a parameterized VDC workload.
The parameterization allows us to tailor the network bandwidth demand
of the VDC workload.
Fourth, the VDC workload should not be problematic by construction,
e.g., an empty datacenter should be able to accommodate the 
most network bandwidth demanding VDC.
Section~\ref{sub:failures} explains several kinds of problems
to be aware of when generating the VDC workload, and 
section~\ref{sub:model} describes a model to avoid these problems.

\subsection{VDC Topologies}
\label{sub:topologies}

\begin{figure}
    \centering
    \includegraphics[width=0.90\linewidth]{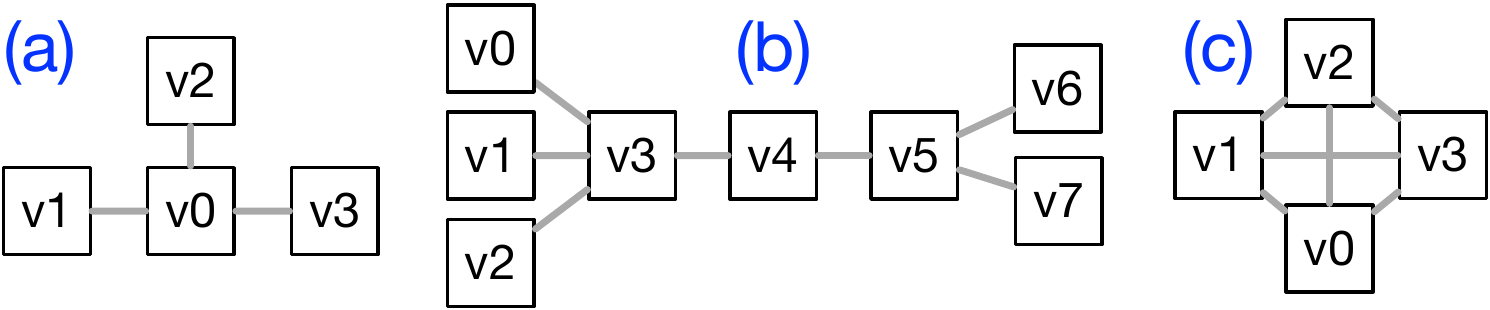}
    \caption{VDC topologies with varying connectivity:
    (a) star topology with sparse connectivity,
    (b) pipeline topology with sparse connectivity, and
    (c) all-to-all topology with dense connectivity.}
    \label{fig:vdc-topologies}
\end{figure}

The design space for VDC topologies ranges from sparse connectivity,
which generates low bandwidth demand, to dense connectivity,
which generated high bandwidth demand.
Fig.~\ref{fig:vdc-topologies} shows three different VDC topologies with
sparse and dense connectivity.
An example VDC with sparse connectivity, as in Fig.~\ref{fig:vdc-topologies}(a),
runs a distributed ML training application in a data-parallel method 
where all worker VMs connect to a
centralized parameter server in a star topology~\cite{parameter-server}.
For example, VDCs used in the existing literature,
such as Yuan et al.~\cite{fmcad13}, SecondNet~\cite{secondnet}, and NetSolver
\cite{netsolver}, have sparse connectivity.
A different example also with sparse connectivity,
as in Fig.~\ref{fig:vdc-topologies}(b),
is a big data application, such as the ETL (Extract-Transform-Load)
Spark pipeline~\cite{aws-etl}.
Here, VMs load data from various sources, process it
with map-reduce-style operations, and
store the results for later processing.
Finally, an example VDC with dense VM connectivity,
as in Fig.~\ref{fig:vdc-topologies}(c),
runs a distributed ML training application in a model-parallel method 
where every VM acts as both a worker and a parameter server.
Here, each VM stores a full copy of the ML model and trains its local
model on its distinct data slice while periodically communicating the
local parameter updates to other VMs.
This method is also called training with
a mirrored strategy~\cite{mirrored-strategy}.

\begin{figure}[t]
    \centering
    \includegraphics[width=0.99\linewidth]{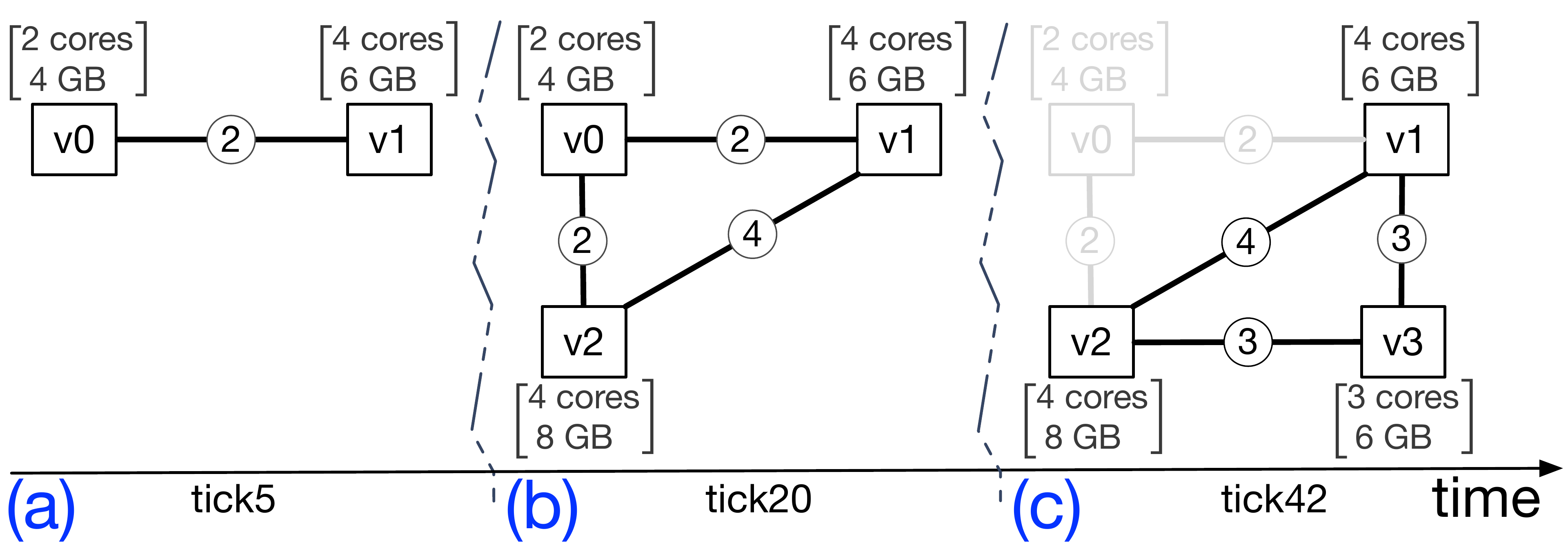}
    \caption{VDC mutation over time: (a) shows VDC creation with two VMs,
    (b) shows VM v2 allocation, and
    (c) shows VM v0 deallocation and VM v3 allocation.
    The VDC will continue with three VMs (v1, v2, v3) after tick 43.}
    \label{fig:vdc-mutation}
\end{figure}

Fig.~\ref{fig:vdc-mutation} shows bandwidth assignment for a sample
VDC with all-to-all topology. It also shows VDC size mutation as
VMs are added to and deleted from the VDC.
Fig.~\ref{fig:vdc-mutation}(a) shows an initial VDC created in
time tick 5 with two VMs: v0 and v1.
The v0-v1 vlink has two ``units of network bandwidth''.\footnote{We will
describe how we derive vlink bandwidths in section
\ref{sub:parameter}.}
We use the term \textit{unit of network bandwidth} throughout our examples
for generality. We convert it to a specific unit, such as 6Mbps,
when we tailor the VDC workload to a specific datacenter
(section~\ref{sub:model}).
The VDC expands to include VM v2 in tick 20,
as shown in Fig.~\ref{fig:vdc-mutation}(b).
This requires allocating v0-v1 and v1-v2 vlinks.
A VDC can also shrink with VM deallocation(s), as shown in tick 42
(Fig.~\ref{fig:vdc-mutation}(c)).
In tick 42, VM v0 is deallocated and VM v3 is allocated.
The VM v0 deallocation requires deallocating the v0-v1 and v0-v2 vlinks.
In the Gridiron technique, VM deletions always precede
VM allocations within a tick so that datacenter resources are first released
before being consumed. 
The VM v3 connects to all other alive VDC VMs, v1 and v2, that requires
allocating v1-v3 and v2-v3 vlinks.
Note that VM deletions in a VDC do not have to adhere to
FIFO (first in first out), LIFO (last in first out), or any other order.
The ordering is inherited from the base workload.

\subsection{Peak VDC Sizes}
\label{sub:vdc-size}

\begin{figure}[t]
    \centering
    \includegraphics[width=0.90\linewidth]{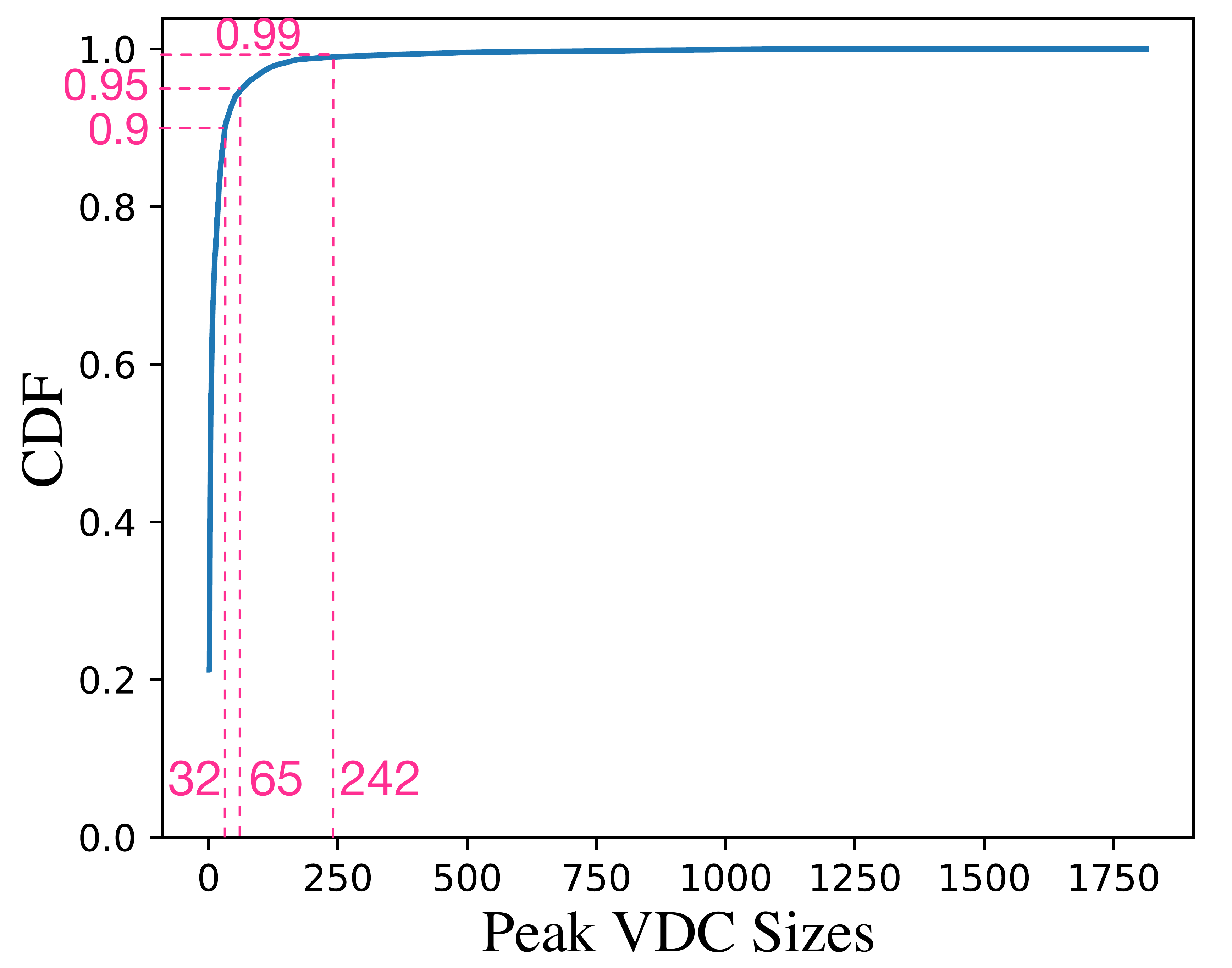}
    \caption[Peak VDC Sizes in the Base Workload]
    {Peak VDC Sizes in the Base Workload.}
    \label{fig:peak-vdc-sizes}
\end{figure}

A VDC reaches its peak size when it has the maximum
number of VMs alive in the same tick.
For example, the VDC in Fig.~\ref{fig:vdc-mutation} reaches peak size
\texttt{P=3} in tick 20 and maintains that size beyond tick 42.
Fig.~\ref{fig:peak-vdc-sizes} shows the distribution of peak VDC
sizes in the base workload. It shows that the 90th percentile (p90)
of the VDCs has 32 VMs and
VDC sizes can reach as many as 1,814 VMs.

However, VDC sizes should be representative of the cloud applications
they encapsulate. For example, the VDC workload will not be realistic if
peak VDC sizes in it are in order of hundreds while there is no such
VDC application that runs at that scale.
The distributed ML training application that we are using
as a motivating example, which significantly benefits from VDC network
bandwidth guarantees~\cite{p3, tictac, bytescheduler} and
is emerging as one of the dominant cloud workloads~\cite{tiresias},
is challenging to run at scale~\cite{facebook-training}.
In particular, with 1,814 VMs.
Therefore, we should be able to cap the peak VDC size when generating
a VDC workload to model a particular cloud application.
In the Gridiron technique, we cap the peak VDC sizes by splitting VDCs.

We split a too-large VDC into multiple VDCs via a rolling-overflow mechanism.
VDCs are processed one-by-one, keeping track of the peak VDC size so far.
If that size exceeds the cap, then subsequent VM allocations to the same
VDC are rolled to a new VDC.
In other words, a VDC will have no more VM create events after it reaches
the cap size.
At the same time, if the base workload VDC has a peak size below the cap,
no capping is applied, and it will be added to the VDC workload as-is,
as a single VDC.
Given that the rolling-overflow mechanism splits a single VDC into
multiple VDCs, the number of VDCs in the capped workload will potentially
exceed the number of VDCs in the base workload.

\subsection{Parameterizing VDC Workload's Network Load}
\label{sub:parameter}

We assign a bandwidth value to each vlink using a
compute-proportional-bandwidth approach.
However, a vlink connects two VMs: Which VM's compute
capacity should be used as the reference point?
In the Gridiron technique, we choose the VM with the smaller capacity,
because doing otherwise introduces a \textit{flooding} effect in practice.
Flooding happens when a VM with more compute capacity sends a higher volume
of network traffic to the VM with less compute capacity,
or the \textit{weaker} VM, to the point where
the weaker VM is no longer able to process the traffic.
The weaker VM drops the subset of packets it cannot process.
Thus, the vlink bandwidth is more realistic if the excess
packets were not sent to begin with.
For example, in Fig.~\ref{fig:vdc-mutation}, the v0-v1 vlink has two units of
network bandwidth as the VM with the weaker processing capacity has two vCPU
cores (VM v0). Similarly, the v1-v2 vlink has four units of network
bandwidth because both VMs that it connects have four vCPU cores.

We generate VDC workloads with varying network demand by parameterizing
each vlink's bandwidth. The key insight in parameterization is that
we can use different ``units'' to represent the unit of bandwidth.
For example, in Fig.~\ref{fig:vdc-mutation}, the v0-v1 vlink gets assigned
2Mbps bandwidth when we use 1Mbps as the unit, and 10Mbps bandwidth
when we use 5Mbps as the unit.
Given that the unit of bandwidth of the vlink is determined by the compute
capacity of the weaker VM (that it connects), we can directly make the
unit as the function of the vCPU cores in the weaker VM.
We call this unit as \textit{bandwidth per core} (bpc).
In our example, the v0-v1 vlink gets assigned 2Mbps bandwidth
when \texttt{bpc=1Mbps}, and 10Mbps bandwidth when \texttt{bpc=5Mbps}
(because the weaker VM has 2 vCPUs).
Thus, we can generate VDC workloads with varying network demand by assigning
different values to the bpc parameter.
For example, the workload with \texttt{bpc=2Mbps} has twice higher
network demand than the workload with \texttt{bpc=1Mbps}.

One can use the bpc parameter to avoid the \textit{over-provisioned}
network pitfall that SecondNet's VDC workload suffered from.
The VDC workload used in the SecondNet~\cite{secondnet} consumed 1/50th
of the bandwidth than that the datacenter provided~\cite{netsolver}.
In other words, the datacenter's network
capacity was 50$\times$ over-provisioned.
Thus, VDCs' network bandwidth requirements were mostly irrelevant during
resource scheduling despite networking being the central focus of the paper.
Although this might be acceptable when a cloud provider is willing to leave a
significant portion of their network bandwidth underutilized,
it seems an unlikely scenario in practice, because cloud providers report the
network to be a computation bottleneck with ToR (top-of-rack) switch uplinks
frequently operating above 80\% utilization~\cite{vl2}.

\subsection{Network-bound VM Allocation Failures}
\label{sub:failures}

Cloud services depend on datacenter capabilities. For example,
cloud providers will not offer a VM flavor with 70 vCPUs if their datacenter
does not have a server with that many cores.
We call this phenomenon \textit{datacenter-level constraints} on cloud services.
We take datacenter-level constraints into account when
generating a VDC workload, because doing otherwise risks generating
VDC workloads that are \textit{problematic by construction}.
In this subsection, we discuss VM allocation failures that surface if
datacenter-level-constraints are not taken into account.
A VM allocation failure happens when a tenant VM allocation request is rejected
because of insufficient residual capacity in the datacenter.

We call a VM allocation failure a network-bound VM allocation failure,
or \textit{network-bound failure} for short, when the VM allocation
request is rejected because of insufficient network bandwidth
in the datacenter.
Avoiding network-bound failures is more complex
than avoiding compute-bound failures because, unlike vCPUs,
inter-VM network bandwidth is not a server-local resource:
it is a cross-device resource.
We identified three specific scenarios that can produce
network-bound failures by construction.
We now elaborate on each scenario and in section~\ref{sub:model},
we present three constraints to enforce during VDC workload generation to
avoid these scenarios.

\begin{figure}[t]
    \centering
    \includegraphics[width=0.70\linewidth]{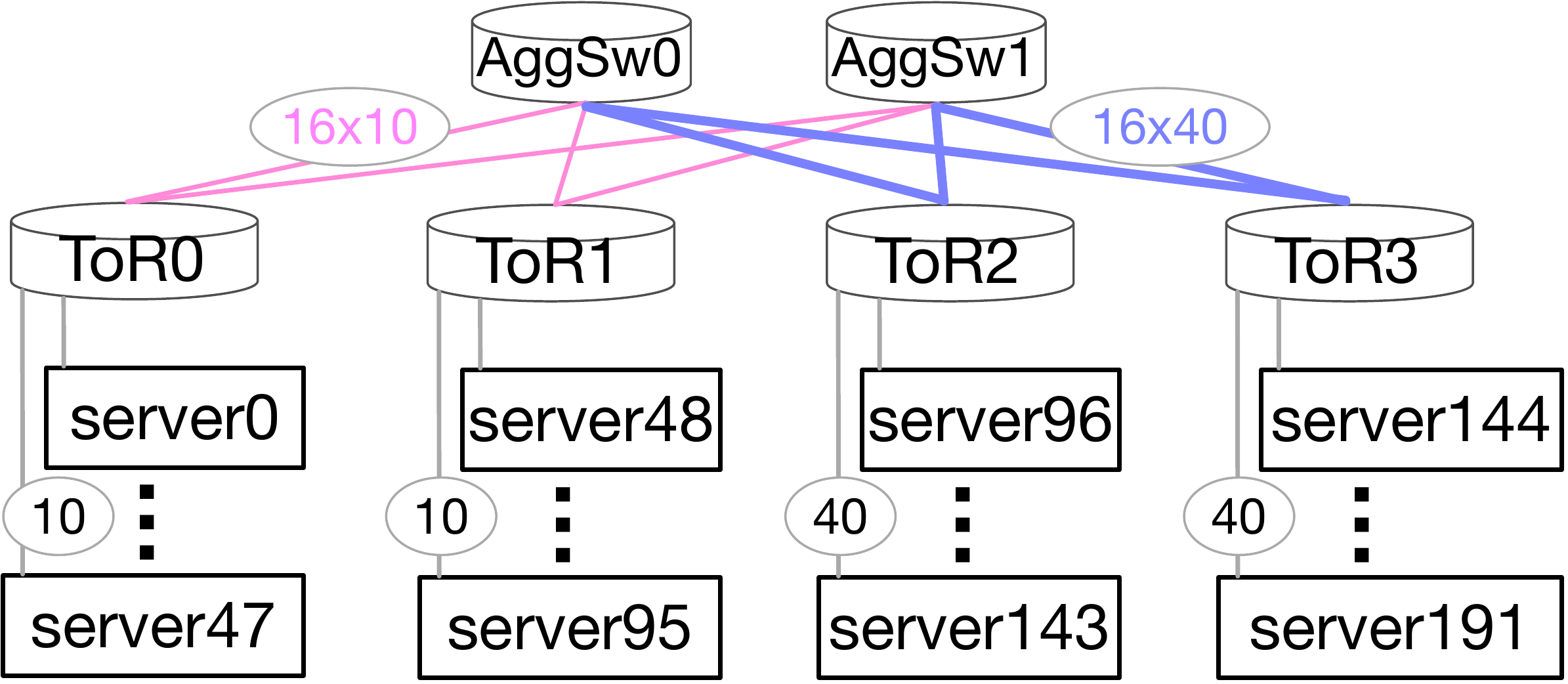}
    \caption{Example Datacenter with Four Racks.
    Here, racks are connected to two aggregation switches.}
    \label{fig:sample-dc}
\end{figure}

The first scenario is allowing excessive bandwidth in a vlink
(\textit{vlink-caused}).
Vlink-caused failures happen when a vlink's bandwidth exceeds the
\textit{aggregate uplink capacity at the most
network-bandwidth-intensive server}, which we call ``fattest server-uplink''.
If servers have two or more uplinks (multi-homed), the fattest server-uplink's
capacity will be equal to a server's aggregate uplink bandwidth.
Vlink-caused failure is analogous to the number of vCPUs exceeding the number of
CPUs at the most compute-intensive server.
For example, for the datacenter shown in Fig.~\ref{fig:sample-dc}, a vlink
with over 40 Gbps is guaranteed to cause a network-bound failure.
(The only exception is if both ends of the vlink are colocated on the same
server.)

We define the ``fattest link'' in terms of ``server uplink''
because of the network multi-path feature, which is commonly used in modern
datacenters~\cite{vl2}. For example,
when two VDC VMs are placed across different racks, such as on server0
and on server144 in Fig.~\ref{fig:sample-dc}, a vlink between these VMs can
span across multiple paths, such as ToR0-AggSw0-ToR3 and ToR0-AggSw1-ToR3,
to pool the bandwidth amount required for this vlink.
In general, the vlink bandwidth should not exceed the bandwidth across
any cut in the network between the servers that host two ends of the vlink.
However, this is non-trivial to compute and depends on the placement of the VMs,
but the fattest server-uplink is always a cut, and therefore an upper bound
on the vlink bandwidth.

\begin{figure}[t]
  \centering
  \includegraphics[width=\linewidth]{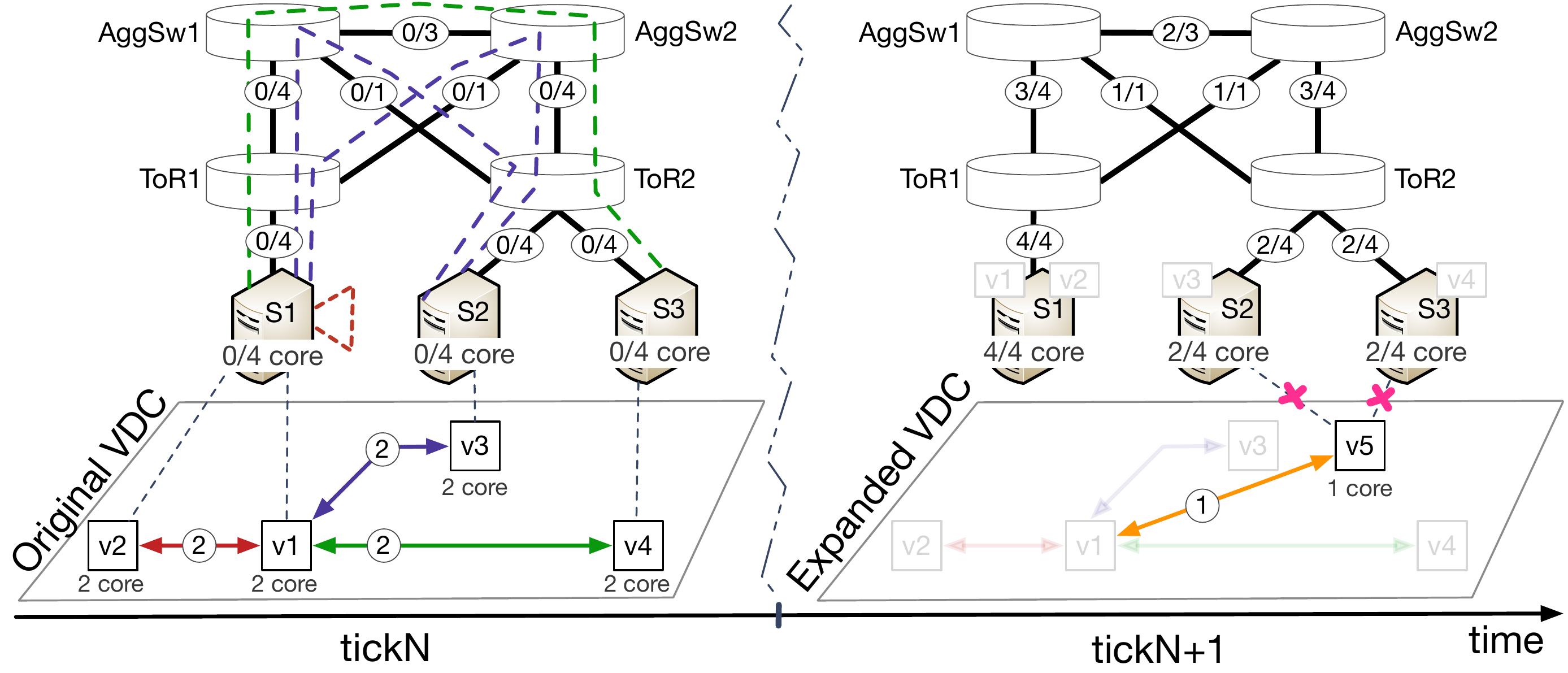}
  \caption{An Example Overpeering-caused Failure.
  The initial VDC allocation succeeds in tick N but the expansion of the VDC
  with VM v5 in tick (N+1) fails because of overpeering of VM v1.
  Datacenter resource capacities are shows as used / original.
  For example, 1/4 in link bandwidth means that ``1'' unit of bandwidth
  is used out of total ``4'' units. We omit VM and server memory capacities
  for brevity. The datacenter is empty in tick N.}
  \label{fig:overpeering}
\end{figure}

The second scenario is allowing excessive VM overpeering\footnote{We
could say ``oversubscription'' instead of ``overpeering'' because
peer VM(s) actually \textit{subscribe} to the already-allocated peers.
However, the term ``oversubscription'' is already used in the cloud's compute
service. For example, saying ``a server is CPU oversubscribed'' means that
``the number of vCPUs VMs are consuming exceed what the host server offers''.}
(\textit{overpeering-caused}).
Overpeering-caused failures happen when a VDCs' already allocated
VM(s) get too many peering requests such that the server hosting
an already-allocated VM becomes network bandwidth bottlenecked.
Fig.~\ref{fig:overpeering} shows an overpeering-caused failure.
In tick N, a tenant requests a VDC with four VMs (v1, v2, v3, v4).
These four VMs get placed on three servers (S1, S2, S3) as shown with the dashed
lines in the left figure. We show VM-to-server assignment with grayed boxes
placed on the servers in tick (N+1).
The tenant requests to expand the original VDC with VM v5 in tick (N+1).
The v5 needs to connect to v1 with 1 unit of network bandwidth.
However, as we can see in the right figure, the only two servers
with sufficient cores and memory to accommodate v5
(S2, S3) do not have sufficient network bandwidth to
connect to S1, which hosts the already allocated peer VM (v1).
Thus, the VDC scheduler has no choice but to fail to allocate v5
due to insufficient network bandwidth.

The overpeering-caused failures happen for the same reason as the
vlink-caused failures: insufficient
network bandwidth at the server uplink level (server-to-ToR switch links).
Similarly to the vlink-caused failures, we do not include failures that happen
due to network bandwidth scarcity in other datacenter network levels,
such as the spine links (because of multi-pathing).
For example, in Fig.~\ref{fig:overpeering},
VM v5 fails because of the S1-ToR1 server uplink.
Otherwise, S2(or S3)-ToR2-AggSw2-AggSw1-ToR1 path does have a 1 unit of network
bandwidth available to accommodate the v1-v5 vlink.

\begin{figure}[t]
  \centering
  \includegraphics[width=0.99\linewidth]{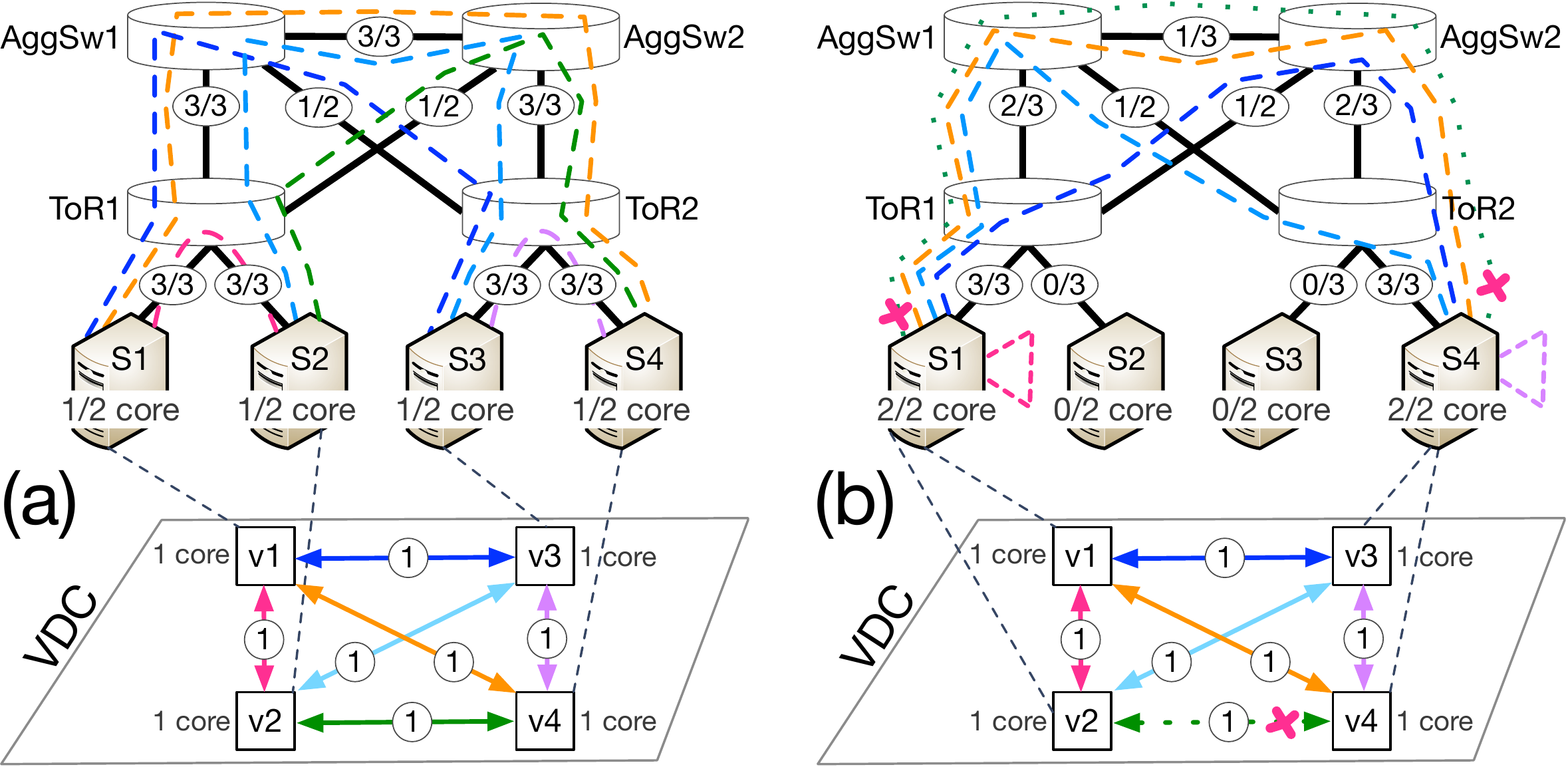}
  \caption{Effect of Colocation on Datacenter Network Bandwidth.
  We show how colocation can over-stress datacenter network:
  (a) shows successful VDC allocation, and
  (b) shows colocation-caused VM allocation failure (v4).
  Datacenter resource capacities are shown as used / original.
  For example, 1/4 in link bandwidth means that ``1'' unit of bandwidth
  is used out of total ``4'' units. The datacenter hosts only this VDC.}
  \label{fig:colocation-caused}
\end{figure}

The third scenario is allowing excessive VM colocation,
which concentrates aggregate vlink bandwidth on a single server
(\textit{colocation-caused}).
We say that two VDC VMs are \textit{colocated} when they are placed on
the same server. Colocation-caused failures happen when the aggregate network
bandwidth to colocated VDC VMs exceeds the bandwidth offered
by the host server. An example is shown in Fig.~\ref{fig:colocation-caused}
where a VDC with four VMs needs to be placed on a datacenter with four servers.
VDC VMs arrive and are placed one-by-one.
Fig.~\ref{fig:colocation-caused}(a) shows successful VDC allocation when
no VMs are colocated.
Fig.~\ref{fig:colocation-caused}(b) shows VM v4 allocation failure
because of colocation.
This VDC consumes the maximal bandwidth on a single
physical link when VDC VMs are equally split across two servers, e.g.,
v1 and v2 VMs are placed on server S1, and two other VMs
are placed on server S4. Colocated VMs do not consume any
datacenter network bandwidth to communicate with each other since they
communicate locally (through the hypervisor), but they do consume datacenter
network bandwidth to communicate with \textit{every} VM placed on the
other server. Thus, there are 2$\times$2=4 vlinks, quadratic in the number
of VMs placed on each server, traversing the path between server S1 and
server S4. However, the S1-ToR1 link can accommodate only 3 vlinks.
Hence, v2-v4 vlink allocation fails, causing VM v4 allocation failure.
Notice that VM v4 in Fig.~\ref{fig:colocation-caused}(b) fails allocation
even though colocation reduces the overall demand on datacenter network
bandwidth (4 units) compared to overall bandwidth without colocation
(6 units in Fig.~\ref{fig:colocation-caused}(a)).
This demonstrates that the colocation-caused failures happen due to demand
concentration, not necessarily demand increase.

In summary, we described three scenarios that can cause network-bound failures
due to datacenter-level constraints.
These scenarios should be taken into account when generating any VDC workload.
Otherwise, the VDC workload can be problematic by construction,
because it may produce vlink-caused, overpeering-caused, and
colocation-caused failures.
Moreover, these scenarios are not exhaustive. There could be other scenarios
that cause network-bound failures. 
However, we believe that these three scenarios are the most relevant ones
to take into account during VDC workload generation.

\subsection{Avoiding Network-bound VM Allocation Failures}
\label{sub:model}

VDC workloads should not be problematic by construction.
We call such workloads \textit{datacenter-aware}.
We present a series of three constraints
for generating datacenter-aware VDC workloads,
which avoid vlink-, overpeering-, and colocation-caused failures.

There are three knobs to control: (1) maximum bandwidth per vlink,
(2) VDC topology, and (3) peak VDC size.
The first knob is similar to the number of vCPUs in a VM flavor.
We can cap vlink capacities to ensure that the highest bandwidth
a vlink offers does not exceed the capacity of the fattest server-uplink.
For example, for the datacenter shown in
Fig.~\ref{fig:sample-dc}, no vlink should offer over 40 Gbps.
Two other knobs, VDC topology and peak VDC size, are related to each other
and can be constrained to avoid overpeering-caused
and colocation-caused failures.

Fig.~\ref{fig:overpeering} shows that in constructing a VDC workload,
we need to budget not only for a VM's current network bandwidth requirements
but also for its \textit{growth potential}.
A VM's growth potential is the difference between the network bandwidth
it consumes at its allocation and how much more bandwidth it can consume
in the future. For example, if a VM is created with only one vlink that has
100Mbps bandwidth but it can create 10 more such vlinks during its lifetime,
this VM's growth potential is 1000Mbps (11*100-100).
A VMs' growth potential is a function of two things:
the topology of its VDC, which determines the number of vlinks the VM can have,
and the peak VDC size, which defines
the maximum number of VMs allowed in a VDC at the same time.
Among VDC topologies with sparse and dense peering
in Fig.~\ref{fig:vdc-topologies},
the dense VDC topology have the highest growth potential
(Fig.~\ref{fig:vdc-topologies}(c)).
This potential can induce overpeering-caused and colocation-caused failures,
because VMs have the highest aggregate bandwidth when they peer with every
other VM in the VDC, i.e., in an all-to-all topology.

We can impose restrictions on the VDC topology to avoid VDCs with dense
connectivity.
For example, we could allow only two vlinks per VM to ensure that all VDCs
have a sparse topology, e.g., a chain-like topology.
However, this would be too restrictive.
As an example, the generated VDC workload could not contain
distributed-ML-training-like applications that have all-to-all connectivity.
Instead, we can just cap the peak VDC size while granting
tenants complete freedom in choosing the VDC topology.
Put differently, we can guard against the overpeering-caused
and colocation-caused failures by controlling only the peak VDC size knob,
and leave the topology knob free by assuming
(the worst case) all-to-all VDC topology.

\begin{figure*}[t]
  \centering
  \includegraphics[width=0.90\linewidth]{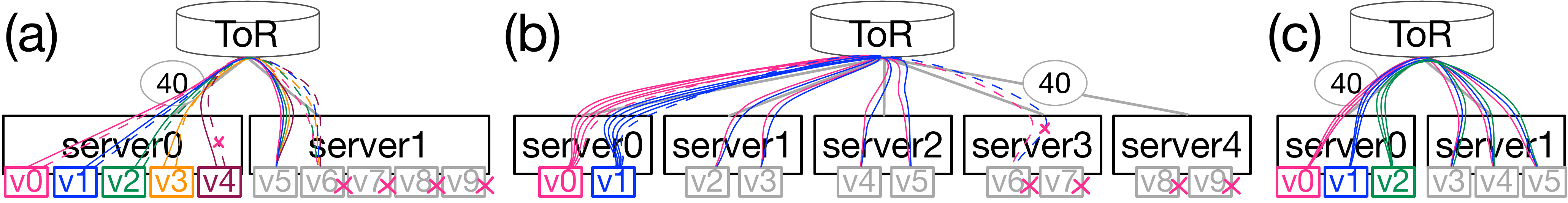}
  \caption{Effect of Colocation on Datacenter Network Bandwidth.
  We show how colocation can create bottleneck(s) in datacenter network.
  VM-to-server placement is shown as VMs placed on the servers:
  (a) and (b) show four colocation-caused VM failures (v6, v7, v8, v9), and
  (c) shows failure-free VDC placement.
  VDCs in all figures have all-to-all connectivity. Figure (b) shows
  connectivity for only VM v0 and VM v1 for brevity.}
  \label{fig:colocation-constraint}
\end{figure*}

For example, consider the 4-rack datacenter topology in Fig.~\ref{fig:sample-dc}
(we generalize this example later):

\begin{enumerate}
  \item We can avoid vlink-caused failures by capping vlink capacities
  to $C=40$ Gbps, i.e., the fattest server-uplink capacity.

  \item Avoiding overpeering-caused failures is about capping VDC size
  such that the aggregate bandwidth of the VM vlinks does not exceed \texttt{C}.
  For the sample datacenter in Fig.~\ref{fig:sample-dc}, we need
  to limit the per-VM aggregate bandwidth to 40 Gbps.
  Assume that the peak VDC size $P = 10$.
  A VDC VM can therefore can have up to $P-1=9$ vlinks. There,
  we need to ensure that the aggregate bandwidth of 9 vlinks
  does not exceed 40 Gbps. Thus, we cap each vlink to have at most
  $B=40/9 \approx 4.4$ Gbps bandwidth to avoid
  overpeering-caused VM allocation failures.

  \item We can prevent colocation-caused failures by avoiding VM
  colocation or by capping the VDC VM \textit{colocation degree}.
  A VDC has colocation degree of $D$ when the largest number of VMs of the VDC
  colocated on a server is $D$.

  For example, imagine the VDC with 10 VMs shown in
  Fig.~\ref{fig:colocation-constraint}(a).
  The VDC has all-to-all topology where each vlink has 4.4 Gbps bandwidth
  ($B$=4.4 Gbps).
  VMs arrive and are placed one-by-one. The first five VMs (v0, v1, v2, v3, v4)
  are colocated on server0, after which server0 becomes compute bound.
  Although server1 has sufficient compute capacity to accommodate the remaining
  five VMs (v5, v6, v7, v8, v9), only VM v5 gets successfully allocated,
  because server0's
  uplink is exhausted after 9 vlinks (five vlinks to connect v5 to the first
  five VMs in server0, and four vlinks to connect v6 to the first four VMs;
  4.4$\times$9$\approx$40)
  and has no bandwidth left for v4-v6 vlink.
  Thus, the last four VMs (v6, v7, v8, v9) fail allocation (when $D=5$).

  Imagine placing this VDC on a five-server datacenter, shown in
  Fig.~\ref{fig:colocation-constraint}(b). Here also, VDC VMs are placed
  one-by-one and each vlink has 4.4 Gbps bandwidth.
  Assume that servers' compute capacity suffices to accommodate only two VMs.
  The first six VMs get successfully allocated, after which server0's
  uplink is exhausted with 9 vlinks (4.4$\times$9$\approx$40)
  and has no bandwidth left for v1-v6 vlink.
  Thus, the last four VMs fail allocation even when $D=2$.
  Therefore, for the datacenter shown in Fig.~\ref{fig:sample-dc}, where
  $C=40$ Gbps, $P=10$, and $B=4.4$ Gbps,
  disabling colocation altogether ($D=1$), is the only way to
  avoid colocation-caused failures.

  At the same time, Fig.~\ref{fig:colocation-constraint}(c) shows that it is
  possible to leave the colocation degree unconstrained
  when $P=6$ (while $C=40$ Gbps and $B=4.4$ Gbps)
  because 40 Gbps server-to-ToR links can accommodate the maximal
  aggregate bandwidth of the smaller, 6-VM VDC ($D=3$).
  These examples in Fig.~\ref{fig:colocation-constraint} demonstrate
  that we can avoid colocation-caused failures by controlling peak VDC size
  (\texttt{P}) as well as maximal colocation degree (\texttt{D}).
\end{enumerate}

Now we generalize our findings from the sample datacenter and propose
a method for generating datacenter-aware VDC workload.
The datacenter-aware VDC workload should satisfy the following three constraints
to avoid all three causes of network-bound failures:
\begin{enumerate}
    \item vlink-caused failures:
    \begin{equation}
    {B \leq C}
    \label{eq:vlink}
    \end{equation}

    where \texttt{B} is the maximal per vlink bandwidth and 
    \texttt{C} is the capacity of the fattest server-uplink.

    \item overpeering-caused failures:
    \begin{equation}
    {B \leq C / (P-1)}
    \label{eq:overpeering}
    \end{equation}

    where \texttt{P} is the peak VDC size.

    \item colocation-caused failures:
    \begin{equation}
    {B \leq C / (P/2)^2}
    \label{eq:colocation}
    \end{equation}
\end{enumerate}

Note that \eqref{eq:overpeering} supersedes~\eqref{eq:vlink}, and
\eqref{eq:colocation} supersedes~\eqref{eq:overpeering}, when $P \geq 2$.
That is, limiting vlink bandwidth (\texttt{B}) to satisfy the constraint in
\eqref{eq:colocation} automatically satisfies the two other constraints.
At the same time, $P \geq 2$ is always true by VDC construction, because
otherwise a VDC can never have more than one VM ($P = 1$).
By definition, a collection of VMs form a VDC only when
there are two or more VMs in the collection that connect over the network.
Thus, in a VDC, $P \geq 2$ always holds and we
can exclusively focus on satisfying~\eqref{eq:colocation}
(avoiding colocation-caused failures).

The method's purpose is not to prevent all possible network-bound failures,
but to bound vlink bandwidths such that the generated VDC workload is
datacenter-aware. The method is useful in generating VDC workloads that are
sufficiently network intensive to evaluate the VDC schedulers,
but are not problematic by construction.

\section{Case Study: Applying the Gridiron Technique to ML Training Application}
\label{sec:ml}

The peak VDC size in the base workload (1,814 VMs) is too large for
ML training applications. Thus, we cap the peak VDC sizes to a realistic size.
We need to take scalability properties of the ML training applications into
account to derive a meaningful VDC size.

A common way to scale distributed ML training is by parallelism.
For example, Goyal et al.\ scale DNN training by increasing the training batch
size and executing data-parallel training across multiple machines/devices
\cite{facebook-training}.
In a VDC, multiple VMs would execute the data-parallel training:
the training data is split across VDC VMs, and each VM runs computation
on a slice of the local data (mini-batch). The result of the computation
on a mini-batch (gradients) is broadcast to all other VDC VMs.
A VM applies gradients from all VDC VMs to its local model,
which is called training the model in batches.
For example, in a VDC with four VMs using batch size 64, each VM will have
mini-batch size 16. More generally, we use the following formula to derive
the number of VMs in a VDC:\\
\centerline{$Number~of~VMs = \frac{Batch~Size}{Mini~Batch~Size}$}

\begin{table*}[t]
    \caption{Common Distributed DNN Training Applications.
    Models are sorted by their size.}
    \label{tbl:ml-workloads}
    \centering\resizebox{0.90\linewidth}{!}{
    \begin{tabular}{p{0.13\linewidth} p{0.10\linewidth} p{0.12\linewidth}
    p{0.09\linewidth} p{0.11\linewidth} p{0.11\linewidth} c}
    \hline
    \textbf{Task} & \raggedright{\textbf{Model Name}} &
    \raggedright{\textbf{Model Size (MB)}} & \raggedright{\textbf{Batch Size}} &
    \raggedright{\textbf{Mini-Batch Size}} & \raggedright{\textbf{Number of VMs}} & \\

    \hline
    \raggedright{Recommendation} &
    \raggedright{DeepLight~\cite{DeepLight-small-scale}}  &
    \raggedright{2319} & \raggedright{$2^{11}$--$2^{13}$} &
    \raggedright{2048} & \raggedright{1--4} & \\

    \raggedright{Translation} & \raggedright{LSTM~\cite{lstm}} &
    \raggedright{1627} & \raggedright{8--64} & \raggedright{8--32} &
    \raggedright{1--8} & \\

    \raggedright{Translation} & \raggedright{BERT~\cite{bert}} &
    \raggedright{1274} & \raggedright{4--256} & \raggedright{4--32} &
    \raggedright{8--64} & \\

    \raggedright{Image classification} &
    \raggedright{VGG19~\cite{VGG19-large-scale}} & \raggedright{548} &
    \raggedright{64--256} & \raggedright{32--256} & \raggedright{1--8} & \\

    \raggedright{Translation} &
    \raggedright{UGATIT~\cite{U-GAT-IT-small-scale}} & \raggedright{511} &
    \raggedright{1--2} & \raggedright{1} & \raggedright{1--2} & \\

    \raggedright{Recommendation} & \raggedright{NCF~\cite{NCF-small-scale}} &
    \raggedright{121} & \raggedright{128--$2^{17}$} &
    \raggedright{128--$2^{14}$} & \raggedright{1--8} & \\

    \raggedright{Object detection} & \raggedright{SSD~\cite{SSD-small-scale}} &
    \raggedright{98} & \raggedright{1--8} & \raggedright{1--8} &
    \raggedright{1--8} & \\

    \raggedright{Image classification} &
    \raggedright{ResNet-50~\cite{resnet50-small-scale}} & \raggedright{87} &
    \raggedright{64--$2^{21}$} & \raggedright{32--8192} &
    \raggedright{8--256} & \\

    \hline
    \end{tabular}}
\end{table*}

%
%
%
%
%
%
%
%

However, higher batch sizes hurt the learning rate, impeding linear 
scalability beyond a certain batch size
\cite{batch-size-scalability, facebook-training, p3}.
An optimal batch size differs by DNN.
Table~\ref{tbl:ml-workloads} lists eight common DNN applications identified
by Sapio et al.~\cite{switchml}.
These applications cover five tasks, out of six total, that were
selected as the representative applications by the MLPerf training benchmarking
organization~\cite{mlperf}, which has the broadest recognition across academia
and industry~\cite{mlcommons}. We surveyed the recent literature to study the
batch sizes and mini-batch sizes used to train these DNNs, which we then
used to derive the VDC size range
\cite{DeepLight-small-scale, switchml, LSTM-small-scale, BERT-large-scale,
VGG19-large-scale, U-GAT-IT-small-scale, NCF-small-scale, SSD-small-scale,
resnet50-small-scale, facebook-training}.\footnote{We report
the studies that use only CPUs and GPUs.
We exclude batch sizes when the model is trained with vendor-specific
accelerators, such as TPUs, because they are uncommon.}

We chose ResNet-50 training as the most common workload.
ResNet-50 is the state-of-the-art model in image classification, and
is the most widely studied model in the literature~\cite{mlperf}.
ResNet-50 achieves linear scalability for batch sizes up to 1024
\cite{facebook-training, p3}. Given that the most commonly used mini-batch size
in the literature is 32,
a VDC to train ResNet-50 can have up to 32 VMs (32x32=1024),
which we round to 30.
We believe that a VDC size of 30 is realistic for modern cloud environments, and
this size also generalizes to models other than ResNet-50 because the workload
analysis study by Jeon et al.\ shows that distributed DNN training
in clusters with up to 16 VMs were already common in 2017~\cite{philli},
and the cluster size has been increasing due to increased DNN model size
\cite{switchml}.\footnote{Note that the Jeon et al.\ study~\cite{philli} is
different from the Resource Central paper~\cite{rc}.
Jeon et al.\ analyzed DNN training workloads deployed on a multi-tenant GPU
cluster in Microsoft during a 75-day period (from 2017.10 to 2017.12).
Their analysis contained 96,260 jobs.}

\begin{figure}[t]
    \centering
    \includegraphics[width=0.70\linewidth]{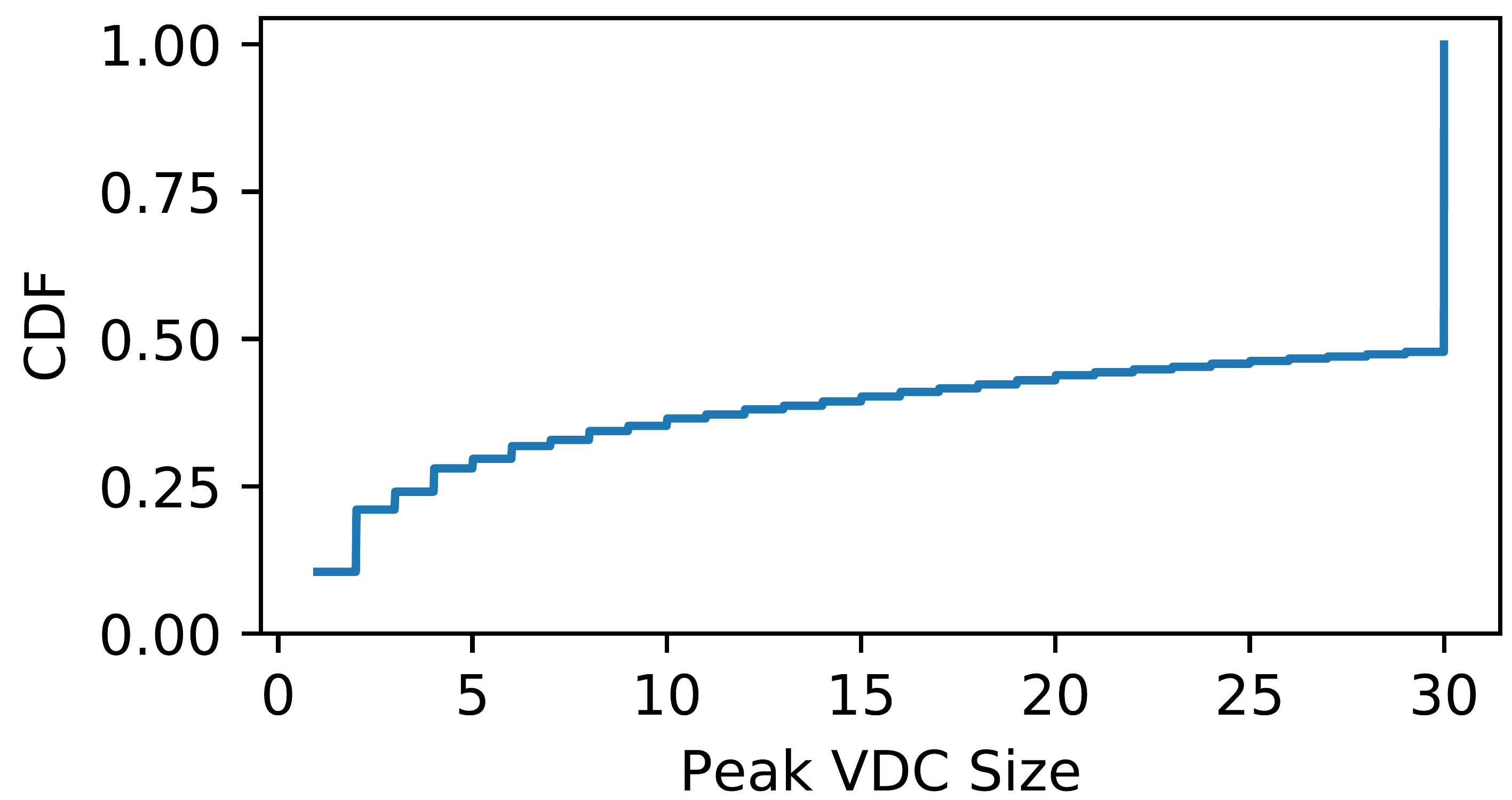}
    \caption{Peak VDC Sizes in the ML Training Workload:
    48\% of the VDCs reach a peak size below 30 and the rest have peak size of
    exactly 30.}\label{fig:peak-ml-vdc-sizes}
\end{figure}

Fig.~\ref{fig:peak-ml-vdc-sizes} shows the peak VDC size distribution
after we cap the peak VDC size at 30.
The VDC workload we construct has $\approx$2$\times$ more VDCs (73,872)
than the base workload (35,870).

Next, to make the generated workload datacenter-aware,
we need to target a specific datacenter.
We demonstrate the vlink bandwidth assignment on the sample 4-rack datacenter
in Fig.~\ref{fig:sample-dc}.
Given the peak VDC size \texttt{P=30} and the fattest server-uplink
capacity in the sample datacenter \texttt{C=40,000} Mbps,
from Fig.~\ref{eq:colocation}:\\
\centerline{$B \leq C / (P/2)^2 = 40,000 / (30/2)^2 \approx 177.78$ Mbps}
\noindent
which means that the VDC workloads will be free from
problematic-by-construction network-bound failures
as long as vlinks' bandwidth do not exceed 177 Mbps.

From the cap on vlink, we can derive a value for bandwidth-per-core
(\texttt{bpc}) parameter. In the base workload, the largest VM has 16 cores.
Therefore, we limit \texttt{bpc$\leq$11Mbps} (177 Mbps / 16 cores) to
ensure that vlinks never exceed 177 Mbps.
We can use different \texttt{bpc} values
to generate VDC workloads with different network demand.
For example, with \texttt{bpc=1Mbps}, the VDC workload consumes
between 5,828 Gbps and 6,581 Gbps.
This variance ($<$10\%) is similar to the variance in
the base workload's CPU footprint. The similarity is by design:
\texttt{bpc} is just a multiplier on the workload's CPU footprint.

\section{Related Work}
\label{sec:relwork}

We divide the related work into two areas. The first area covers existing
techniques for VDC workload generation. We explain how the Gridiron technique
differs from the existing techniques.
The second area covers publicly available traces from production clouds.
We explain our rationale for choosing the Resource Central dataset
\cite{dataset} for VDC workload generation.

\subsection{Existing Techniques for Generating VDC Workloads}

To the best of our knowledge, no existing work uses production
cloud workload traces to construct a VDC workload.
Instead, all existing work use a randomization-based approach for
generating VDCs and their lifetimes~\cite{oktopus, fmcad13, secondnet,
sun2017efficient, yang2015towards, greenhead}.
For example, Amokrane et al.~\cite{greenhead} generate VDCs with 5--200 VMs,
where a pair of VMs connect with a probability 0.5 with a bandwidth demand
uniformly distributed between 10 and 50 Mbps.
The VDC allocation requests arrive according to a Poisson process,
and VDC lifetimes follow an exponential distribution.
We are the first to generate a VDC workload from production cloud traces
where VDC sizes and VDC allocation requests are directly derived from the
production traces.

\subsection{Rational for Choosing the Resource Central Dataset}

Table~\ref{tbl:workloads} compares Resource Central dataset~\cite{dataset}
with other datasets released by public cloud providers.
The Resource Central dataset is well-suited for VDC workload
generation because, a VDC is a collection of \textit{VMs}, and
the Resource Central dataset contains VMs.
Moreover, the dataset describes VMs with their absolute number of CPU cores,
which is essential for the compute-proportional-bandwidth generation approach.
Lastly, VMs are grouped into ``deployments'' that we can use 
as a proxy for establishing VDC VM membership (section~\ref{sec:base}).

Bergsma et al. propose an approach based on Recurrent Neural Networks (RNN)
to generate realistic cloud (VM) workloads~\cite{workload-generation}.
They also use the Resource Central dataset~\cite{dataset} to validate
properties of their generated workloads.
It is possible to use the RNN-based workload as the base VM workload
and apply the Gridiron technique to generate realistic VDC workloads.
We leave this to future work.

\begin{table*}[t]
    \caption[Production Cloud Workloads]{Publicly Available Production Cloud
    Workloads.}\label{tbl:workloads}
    \centering\resizebox{0.90\linewidth}{!}{
    \begin{tabular}{l p{0.07\linewidth} p{0.07\linewidth}
    p{0.04\linewidth} p{0.11\linewidth}
    p{0.05\linewidth} p{0.025\linewidth} p{0.04\linewidth} c}
    \hline
    Workload & \centering{Virtualization} & \centering{CPU cores}
    & \centering{Grouping} & \centering{Volume (million)}
    & \centering{Duration} & \centering{Year} & \centering{Provider} & \\

    \hline
    \textbf{Resource Central}~\cite{rc} & \centering{VM} & Absolute & \centering{\checkmark} & \centering{2} & \centering{30 days} & \centering{2017} & \centering{Azure} & \\

    Resource Central V2~\cite{azure-dataset-v2} & \centering{VM} & Buckets & \centering{\checkmark} & \centering{2.6} & \centering{30 days} & \centering{2019} & \centering{Azure} & \\

    Protean~\cite{protean} & \centering{VM} & Normalized & & \centering{5.5} & \centering{14 days} & \centering{2020} & \centering{Azure} & \\

    Serverless~\cite{serverless} & \centering{Functions} & & \centering{\checkmark} & \centering{0.6} & \centering{14 days} & \centering{2019} & \centering{Azure} & \\

    Borg 2011~\cite{borg2011-analysis} & \centering{Container} & Normalized & \centering{\checkmark} & \centering{25.4} & \centering{30 days} & \centering{2011} & \centering{Google} & \\

    Borg 2019~\cite{borg2019} & \centering{Container} & Normalized & \centering{\checkmark} & \centering{$\sim$732} & \centering{31 days} & \centering{2019} & \centering{Google} & \\

    Alibaba 2017~\cite{alibaba-2017} & \centering{Container} & Absolute & \centering{\checkmark} & \centering{0.09} & \centering{0.5 day} & \centering{2017} & \centering{Alibaba} & \\

    Alibaba 2018~\cite{alibaba-2018} & \centering{Container} & Absolute & \centering{\checkmark} & \centering{14.3} & \centering{8 days} & \centering{2018} & \centering{Alibaba} & \\

    \hline
    \end{tabular}}
\end{table*}

The same Azure team released version 2 (V2) of the Resource Central
dataset in 2019~\cite{azure-dataset-v2}.
Although this V2 dataset has $\approx$33\% more VMs and is more recent,
it is ill-suited as the VM workload source
because it does not describe VMs' CPU cores in absolute terms.
In the V2 dataset, the authors
anonymized the number of VM CPU cores by
grouping the number of cores into six buckets.\footnote{The first bucket
contains all VMs with $<2$ cores,
or bucket1$<2$ cores for short,
2 cores $\leq$ bucket2 $<4$ cores,
4 cores $\leq$ bucket3 $<8$ cores,
8 cores $\leq$ bucket4 $<12$ cores,
12 cores $\leq$ bucket5 $<24$ cores,
bucket6 $\geq 24$ cores.}
Although it is possible to generate VDCs' inter-VM network bandwidth
requirements in the same granularity as bucket compute capacities,
we believe that using
the more precise CPU core numbers, as in the Resource Central 2017 dataset
\cite{dataset}, allows us to construct a more realistic VDC workload.

The Azure cloud team also released a larger dataset, containing 5.5 million 
VM allocations and deallocations, collected over a period of 14 days.
This dataset is described in and released with the Protean paper
\cite{protean}.
There are two disadvantages to using this dataset as the base VM workload.
First, VM CPU cores are not given in absolute numbers.
They are normalized to the server that the VM is placed on. For example,
if the VM requested 4 cores and the server it is placed on has 40 cores,
the VM will be recorded as having 0.1 cores.
Reverse engineering this dataset to extract absolute cores cannot be
precise because the dataset includes multiple servers configurations
and these configurations are not released.
Moreover, VMs in the Protean dataset do not have grouping, i.e.,
all VMs are solo-VMs. We cannot construct a VDC from solo-VMs,
as discussed in section~\ref{sec:base}.
Therefore, the Protean dataset is also ill-suited for using as the
base VM workload.

The bottom five datasets in Table~\ref{tbl:workloads} do not use VMs as the
virtualization unit. They use functions,
e.g., Azure Functions~\cite{azure-functions},
or containers, e.g., the Azure Container Service~\cite{azure-containers}.
Unfortunately, VM lifetimes are radically different from
containers and function lifetimes: 
VM lifetimes are on the order of \textit{hours}, container lifetimes are
on the order of \textit{minutes}, and function lifetimes are on the order of
\textit{seconds}.\footnote{
More precisely, 50\% of functions run for less than 3 seconds,
or p50=3 seconds for short, and p90=60 seconds~\cite{serverless}.
Lifetime for containers are: p50=50 seconds and
p90=1000 seconds~\cite{alibaba-2018}.
VMs lifetimes are the longest: p50=900 seconds (15 mins) and
p90=86,400 seconds (24 hours)~\cite{rc}.}
Thus, a container trace or function trace is not an ideal starting point
for a VM-based VDC workload.


\section{Conclusions}
\label{sec:conclusion}

We described the Gridiron technique to construct a realistic
VDC workload. We used a VM trace from the Azure production cloud
as the base workload and augmented its VMs with network bandwidth requirements.
We used the notion of ``deployment'' that is present in the Azure trace
to derive a VM's VDC membership.
We considered various VDC topologies and selected all-to-all connectivity.

We proposed the compute-proportional-bandwidth approach
to develop a parameterized VDC workload generation mechanism.
This mechanism allows us to adapt VDC workloads to different datacenters.
For example, we can use this mechanism to scale up a workload's bandwidth
requirements to make sure that the datacenter network is not
over-provisioned to the point that hinders evaluation of
VDC scheduling algorithms' efficacy.

We studied the datacenter-level constraints for VDC workloads.
We described problematic-by-construction network-bound VM allocation
failures and proposed a model to avoid them.
Finally, we applied the Gridiron technique to generate a VDC workload that
captures the characteristics of distributed ML training applications.
We capped the peak VDC size at 30 VMs
to capture the scalability limitations of the
distributed ML training application.
The capping allows us to generate more realistic VDC workloads.

\bibliographystyle{IEEEtran}
\bibliography{iiswc-paper}

\begin{thebibliography}{10}
\providecommand{\url}[1]{#1}
\csname url@samestyle\endcsname
\providecommand{\newblock}{\relax}
\providecommand{\bibinfo}[2]{#2}
\providecommand{\BIBentrySTDinterwordspacing}{\spaceskip=0pt\relax}
\providecommand{\BIBentryALTinterwordstretchfactor}{4}
\providecommand{\BIBentryALTinterwordspacing}{\spaceskip=\fontdimen2\font plus
\BIBentryALTinterwordstretchfactor\fontdimen3\font minus
  \fontdimen4\font\relax}
\providecommand{\BIBforeignlanguage}[2]{{%
\expandafter\ifx\csname l@#1\endcsname\relax
\typeout{** WARNING: IEEEtran.bst: No hyphenation pattern has been}%
\typeout{** loaded for the language `#1'. Using the pattern for}%
\typeout{** the default language instead.}%
\else
\language=\csname l@#1\endcsname
\fi
#2}}
\providecommand{\BIBdecl}{\relax}
\BIBdecl

\bibitem{secondnet}
\BIBentryALTinterwordspacing
C.~Guo, G.~Lu, H.~J. Wang, S.~Yang, C.~Kong, P.~Sun, W.~Wu, and Y.~Zhang,
  ``{SecondNet: A Data Center Network Virtualization Architecture with
  Bandwidth Guarantees},'' in \emph{{Proceedings of the 6th International
  COnference on emerging Networking EXperiments and Technologies}}, ser.
  Co-NEXT '10.\hskip 1em plus 0.5em minus 0.4em\relax Association for Computing
  Machinery, 2010. [Online]. Available:
  \url{https://doi.org/10.1145/1921168.1921188}
\BIBentrySTDinterwordspacing

\bibitem{oktopus}
\BIBentryALTinterwordspacing
H.~Ballani, P.~Costa, T.~Karagiannis, and A.~Rowstron, ``{Towards Predictable
  Datacenter Networks},'' in \emph{Proceedings of the ACM SIGCOMM 2011
  Conference}, ser. SIGCOMM '11.\hskip 1em plus 0.5em minus 0.4em\relax
  Association for Computing Machinery, 2011, p. 242–253. [Online]. Available:
  \url{https://doi.org/10.1145/2018436.2018465}
\BIBentrySTDinterwordspacing

\bibitem{netsolver}
\BIBentryALTinterwordspacing
S.~Bayless, N.~Kodirov, S.~M. Iqbal, I.~Beschastnikh, H.~H. Hoos, and A.~J. Hu,
  ``{Scalable constraint-based virtual data center allocation},''
  \emph{Artificial Intelligence}, vol. 278, 2020. [Online]. Available:
  \url{https://doi.org/10.1016/j.artint.2019.103196}
\BIBentrySTDinterwordspacing

\bibitem{rc}
\BIBentryALTinterwordspacing
E.~Cortez, A.~Bonde, A.~Muzio, M.~Russinovich, M.~Fontoura, and R.~Bianchini,
  ``{Resource Central: Understanding and Predicting Workloads for Improved
  Resource Management in Large Cloud Platforms},'' in \emph{Proceedings of the
  26th Symposium on Operating Systems Principles}, ser. SOSP '17.\hskip 1em
  plus 0.5em minus 0.4em\relax {USENIX} Association, 2017, pp. 153--167.
  [Online]. Available: \url{https://doi.org/10.1145/3132747.3132772}
\BIBentrySTDinterwordspacing

\bibitem{ec2-beta}
\BIBentryALTinterwordspacing
J.~Barr, ``{Amazon EC2 Beta},'' 2006. [Online]. Available:
  \url{https://aws.amazon.com/blogs/aws/amazon\_ec2\_beta}
\BIBentrySTDinterwordspacing

\bibitem{p3}
A.~Jayarajan, J.~Wei, G.~Gibson, A.~Fedorova, and G.~Pekhimenko,
  ``{Priority-based Parameter Propagation for Distributed DNN Training},''
  2019, https://arxiv.org/abs/1905.03960.

\bibitem{tictac}
S.~H. Hashemi, S.~A. Jyothi, and R.~H. Campbell, ``{TicTac: Accelerating
  Distributed Deep Learning with Communication Scheduling},'' 2018,
  https://arxiv.org/abs/1803.03288.

\bibitem{bytescheduler}
\BIBentryALTinterwordspacing
Y.~Peng, Y.~Zhu, Y.~Chen, Y.~Bao, B.~Yi, C.~Lan, C.~Wu, and C.~Guo, ``{A
  Generic Communication Scheduler for Distributed DNN Training Acceleration},''
  in \emph{Proceedings of the 27th ACM Symposium on Operating Systems
  Principles}, ser. SOSP '19.\hskip 1em plus 0.5em minus 0.4em\relax ACM, 2019,
  p. 16–29. [Online]. Available:
  \url{https://doi.org/10.1145/3341301.3359642}
\BIBentrySTDinterwordspacing

\bibitem{tiresias}
\BIBentryALTinterwordspacing
J.~Gu, M.~Chowdhury, K.~G. Shin, Y.~Zhu, M.~Jeon, J.~Qian, H.~Liu, and C.~Guo,
  ``{Tiresias: A GPU Cluster Manager for Distributed Deep Learning},'' in
  \emph{16th USENIX Symposium on Networked Systems Design and Implementation
  (NSDI 19)}.\hskip 1em plus 0.5em minus 0.4em\relax USENIX Association, 2019,
  pp. 485--500. [Online]. Available:
  \url{https://www.usenix.org/conference/nsdi19/presentation/gu}
\BIBentrySTDinterwordspacing

\bibitem{parallel-computing}
Wikipedia, ``{Parallel computing},'' 2020,
  https://en.wikipedia.org/wiki/Parallel\_computing.

\bibitem{amdalh-2nd-law}
\BIBentryALTinterwordspacing
G.~M. Amdahl, ``{Computer Architecture and Amdahl's Law},'' \emph{Computer},
  vol.~46, no.~12, pp. 38--46, 2013. [Online]. Available:
  \url{https://doi.org/10.1109/MC.2013.418}
\BIBentrySTDinterwordspacing

\bibitem{balanced-system}
\BIBentryALTinterwordspacing
F.~{Liang}, C.~{Feng}, X.~{Lu}, and Z.~{Xu}, ``{Performance Characterization of
  Hadoop and Data MPI Based on Amdahl's Second Law},'' in \emph{The 9th IEEE
  International Conference on Networking, Architecture, and Storage}, 2014, pp.
  207--215. [Online]. Available: \url{https://doi.org/10.1109/NAS.2014.39}
\BIBentrySTDinterwordspacing

\bibitem{gcp-network}
C.~McAnlis, ``{5 steps to better GCP network performance},'' 2017,
  https://cloud.google.com/blog/products/gcp/5-steps-to-better-gcp-network-performance.

\bibitem{dataset}
M.~Azure, ``{Microsoft Azure VM Traces (AzurePublicDatasetV1)},'' 2017,
  https://github.com/Azure/AzurePublicDataset.

\bibitem{timestamp}
N.~Kodirov, ``{VM creates with invalid timestamps},'' 2019,
  https://github.com/Azure/AzurePublicDataset/issues/4.

\bibitem{parameter-server}
Tensorflow, ``{Parameter Server Training},'' 2020,
  \url{https://www.tensorflow.org/tutorials/distribute/parameter_server_training}.

\bibitem{fmcad13}
\BIBentryALTinterwordspacing
Y.~Yuan, A.~Wang, R.~Alur, and B.~T. Loo, ``{On the Feasibility of Automation
  for Bandwidth Allocation Problems in Data Centers},'' in \emph{{Formal
  Methods in Computer-Aided Design, 2013}}, ser. FMCAD' 13, 2013, pp. 42--45.
  [Online]. Available: \url{https://doi.org/10.1109/FMCAD.2013.6679389}
\BIBentrySTDinterwordspacing

\bibitem{aws-etl}
Coursera, ``{How Coursera Manages Large-Scale ETL using AWS Data Pipeline and
  Dataduct},'' 2015,
  https://aws.amazon.com/blogs/big-data/how-coursera-manages-large-scale-etl-using-aws-data-pipeline-and-dataduct.

\bibitem{mirrored-strategy}
\BIBentryALTinterwordspacing
Tensorflow, ``{Distributed training with TensorFlow},'' 2020. [Online].
  Available:
  \url{https://www.tensorflow.org/guide/distributed\_training\#mirroredstrategy}
\BIBentrySTDinterwordspacing

\bibitem{facebook-training}
P.~Goyal, P.~Dollár, R.~Girshick, P.~Noordhuis, L.~Wesolowski, A.~Kyrola,
  A.~Tulloch, Y.~Jia, and K.~He, ``{Accurate, Large Minibatch SGD: Training
  ImageNet in 1 Hour},'' 2018, https://arxiv.org/abs/1706.02677.

\bibitem{vl2}
\BIBentryALTinterwordspacing
A.~Greenberg, J.~R. Hamilton, N.~Jain, S.~Kandula, C.~Kim, P.~Lahiri, D.~A.
  Maltz, P.~Patel, and S.~Sengupta, ``{VL2: A Scalable and Flexible Data Center
  Network},'' in \emph{Proceedings of the ACM SIGCOMM 2009 Conference on Data
  Communication}, ser. SIGCOMM '09.\hskip 1em plus 0.5em minus 0.4em\relax
  Association for Computing Machinery, 2009, p. 51–62. [Online]. Available:
  \url{https://doi.org/10.1145/1592568.1592576}
\BIBentrySTDinterwordspacing

\bibitem{DeepLight-small-scale}
W.~Deng, J.~Pan, T.~Zhou, D.~Kong, A.~Flores, and G.~Lin, ``{DeepLight: Deep
  Lightweight Feature Interactions for Accelerating CTR Predictions in Ad
  Serving},'' 2021, https://arxiv.org/abs/2002.06987.

\bibitem{lstm}
\BIBentryALTinterwordspacing
S.~Hochreiter and J.~Schmidhuber, ``{Long Short-Term Memory},'' \emph{Neural
  Comput.}, vol.~9, no.~8, p. 1735–1780, nov 1997. [Online]. Available:
  \url{https://doi.org/10.1162/neco.1997.9.8.1735}
\BIBentrySTDinterwordspacing

\bibitem{bert}
J.~Devlin, M.-W. Chang, K.~Lee, and K.~Toutanova, ``{BERT: Pre-training of Deep
  Bidirectional Transformers for Language Understanding},'' 2019,
  https://arxiv.org/abs/1810.04805.

\bibitem{VGG19-large-scale}
K.~Simonyan and A.~Zisserman, ``{Very Deep Convolutional Networks for
  Large-Scale Image Recognition},'' 2015, https://arxiv.org/abs/1409.1556.

\bibitem{U-GAT-IT-small-scale}
J.~Kim, M.~Kim, H.~Kang, and K.~Lee, ``{U-GAT-IT: Unsupervised Generative
  Attentional Networks with Adaptive Layer-Instance Normalization for
  Image-to-Image Translation},'' 2020, https://arxiv.org/abs/1907.10830.

\bibitem{NCF-small-scale}
X.~He, L.~Liao, H.~Zhang, L.~Nie, X.~Hu, and T.-S. Chua, ``{Neural
  Collaborative Filtering},'' 2017, https://arxiv.org/abs/1708.05031.

\bibitem{SSD-small-scale}
W.~Liu, D.~Anguelov, D.~Erhan, C.~Szegedy, S.~Reed, C.-Y. Fu, and A.~C. Berg,
  ``{SSD: Single Shot MultiBox Detector},'' \emph{Lecture Notes in Computer
  Science}, p. 21–37, 2016,
  \url{http://dx.doi.org/10.1007/978-3-319-46448-0_2}.

\bibitem{resnet50-small-scale}
K.~He, X.~Zhang, S.~Ren, and J.~Sun, ``{Deep Residual Learning for Image
  Recognition},'' 2015, https://arxiv.org/abs/1512.03385.

\bibitem{batch-size-scalability}
N.~S. Keskar, D.~Mudigere, J.~Nocedal, M.~Smelyanskiy, and P.~T.~P. Tang, ``{On
  Large-Batch Training for Deep Learning: Generalization Gap and Sharp
  Minima},'' 2017, https://arxiv.org/abs/1609.04836.

\bibitem{switchml}
\BIBentryALTinterwordspacing
A.~Sapio, M.~Canini, C.-Y. Ho, J.~Nelson, P.~Kalnis, C.~Kim, A.~Krishnamurthy,
  M.~Moshref, D.~Ports, and P.~Richtarik, ``{Scaling Distributed Machine
  Learning with In-Network Aggregation},'' in \emph{18th {USENIX} Symposium on
  Networked Systems Design and Implementation ({NSDI} 21)}.\hskip 1em plus
  0.5em minus 0.4em\relax {USENIX} Association, 2021, pp. 785--808. [Online].
  Available: \url{https://www.usenix.org/conference/nsdi21/presentation/sapio}
\BIBentrySTDinterwordspacing

\bibitem{mlperf}
\BIBentryALTinterwordspacing
P.~Mattson, C.~Cheng, G.~Diamos, C.~Coleman, P.~Micikevicius, D.~Patterson,
  H.~Tang, G.-Y. Wei, P.~Bailis, V.~Bittorf, D.~Brooks, D.~Chen, D.~Dutta,
  U.~Gupta, K.~Hazelwood, A.~Hock, X.~Huang, D.~Kang, D.~Kanter, N.~Kumar,
  J.~Liao, D.~Narayanan, T.~Oguntebi, G.~Pekhimenko, L.~Pentecost,
  V.~Janapa~Reddi, T.~Robie, T.~St~John, C.-J. Wu, L.~Xu, C.~Young, and
  M.~Zaharia, ``{MLPerf Training Benchmark},'' in \emph{Proceedings of Machine
  Learning and Systems}, vol.~2, 2020, pp. 336--349. [Online]. Available:
  \url{https://proceedings.mlsys.org/paper/2020/file/02522a2b2726fb0a03bb19f2d8d9524d-Paper.pdf}
\BIBentrySTDinterwordspacing

\bibitem{mlcommons}
MLCommons, ``{MLCommons},'' 2021, https://mlcommons.org.

\bibitem{LSTM-small-scale}
\BIBentryALTinterwordspacing
K.~Chen and Q.~Huo, ``{Training Deep Bidirectional LSTM Acoustic Model for
  LVCSR by a Context-Sensitive-Chunk BPTT Approach},'' vol.~24, no.~7, p.
  1185–1193, 2016. [Online]. Available:
  \url{https://dl.acm.org/doi/10.5555/2992803.2992806}
\BIBentrySTDinterwordspacing

\bibitem{BERT-large-scale}
N.~AI, ``{BERT Meets GPUs},'' 2019,
  \url{https://medium.com/future-vision/bert-meets-gpus-403d3fbed848}.

\bibitem{philli}
\BIBentryALTinterwordspacing
M.~Jeon, S.~Venkataraman, A.~Phanishayee, J.~Qian, W.~Xiao, and F.~Yang,
  ``{Analysis of Large-Scale Multi-Tenant GPU Clusters for DNN Training
  Workloads},'' in \emph{2019 {USENIX} Annual Technical Conference ({USENIX}
  {ATC} 19)}.\hskip 1em plus 0.5em minus 0.4em\relax Renton, WA: {USENIX}
  Association, 2019, pp. 947--960. [Online]. Available:
  \url{https://www.usenix.org/conference/atc19/presentation/jeon}
\BIBentrySTDinterwordspacing

\bibitem{sun2017efficient}
\BIBentryALTinterwordspacing
G.~Sun, D.~Liao, S.~Bu, H.~Yu, Z.~Sun, and V.~Chang, ``{The Efficient Framework
  and Algorithm for Provisioning Evolving VDC in Federated Data Centers},''
  \emph{Future Generation Computer Systems}, vol.~73, pp. 79--89, 2017.
  [Online]. Available: \url{https://doi.org/10.1016/j.future.2016.12.019}
\BIBentrySTDinterwordspacing

\bibitem{yang2015towards}
\BIBentryALTinterwordspacing
Y.~Yang, X.~Chang, J.~Liu, and L.~Li, ``{Towards Robust Green Virtual Cloud
  Data Center Provisioning},'' \emph{IEEE Transactions on Cloud Computing},
  vol.~5, no.~2, pp. 168--181, 2015. [Online]. Available:
  \url{https://doi.org/10.1016/j.future.2016.12.019}
\BIBentrySTDinterwordspacing

\bibitem{greenhead}
\BIBentryALTinterwordspacing
A.~Amokrane, M.~F. Zhani, R.~Langar, R.~Boutaba, and G.~Pujolle, ``{Greenhead:
  Virtual Data Center Embedding across Distributed Infrastructures},''
  \emph{IEEE transactions on cloud computing}, vol.~1, no.~1, pp. 36--49, 2013.
  [Online]. Available: \url{https://doi.org/10.1109/TCC.2013.5}
\BIBentrySTDinterwordspacing

\bibitem{workload-generation}
S.~Bergsma, T.~Zeyl, A.~Senderovich, and J.~C. Beck, ``{Generating Complex,
  Realistic Cloud Workloads Using Recurrent Neural Networks},'' in
  \emph{Proceedings of the ACM SIGOPS 28th Symposium on Operating Systems
  Principles}, ser. SOSP '21.\hskip 1em plus 0.5em minus 0.4em\relax
  Association for Computing Machinery, 2021, p. 376–391,
  \url{https://doi.org/10.1145/3477132.3483590}.

\bibitem{azure-dataset-v2}
\BIBentryALTinterwordspacing
Azure, ``{AzurePublicDatasetV2},'' 2019. [Online]. Available:
  \url{\url{https://github.com/Azure/AzurePublicDataset/blob/master/AzurePublicDatasetV2.md}}
\BIBentrySTDinterwordspacing

\bibitem{protean}
\BIBentryALTinterwordspacing
O.~Hadary, L.~Marshall, I.~Menache, A.~Pan, E.~E. Greeff, D.~Dion, S.~Dorminey,
  S.~Joshi, Y.~Chen, M.~Russinovich, and T.~Moscibroda, ``{Protean: VM
  Allocation Service at Scale},'' in \emph{14th {USENIX} Symposium on Operating
  Systems Design and Implementation ({OSDI} 20)}.\hskip 1em plus 0.5em minus
  0.4em\relax {USENIX} Association, 2020, pp. 845--861. [Online]. Available:
  \url{https://www.usenix.org/conference/osdi20/presentation/hadary}
\BIBentrySTDinterwordspacing

\bibitem{serverless}
\BIBentryALTinterwordspacing
M.~Shahrad, R.~Fonseca, I.~Goiri, G.~Chaudhry, P.~Batum, J.~Cooke, E.~Laureano,
  C.~Tresness, M.~Russinovich, and R.~Bianchini, ``{Serverless in the Wild:
  Characterizing and Optimizing the Serverless Workload at a Large Cloud
  Provider},'' in \emph{2020 {USENIX} Annual Technical Conference}.\hskip 1em
  plus 0.5em minus 0.4em\relax {USENIX} Association, 2020, pp. 205--218.
  [Online]. Available:
  \url{https://www.usenix.org/conference/atc20/presentation/shahrad}
\BIBentrySTDinterwordspacing

\bibitem{borg2011-analysis}
\BIBentryALTinterwordspacing
C.~Reiss, A.~Tumanov, G.~R. Ganger, R.~H. Katz, and M.~A. Kozuch,
  ``{Heterogeneity and Dynamicity of Clouds at Scale: Google Trace Analysis},''
  in \emph{Proceedings of the Third ACM Symposium on Cloud Computing}, ser.
  SoCC '12.\hskip 1em plus 0.5em minus 0.4em\relax Association for Computing
  Machinery, 2012. [Online]. Available:
  \url{https://doi.org/10.1145/2391229.2391236}
\BIBentrySTDinterwordspacing

\bibitem{borg2019}
\BIBentryALTinterwordspacing
M.~Tirmazi, A.~Barker, N.~Deng, M.~E. Haque, Z.~G. Qin, S.~Hand,
  M.~Harchol-Balter, and J.~Wilkes, ``{Borg: the Next Generation},'' in
  \emph{Proceedings of the Fifteenth European Conference on Computer Systems
  (EuroSys'20)}.\hskip 1em plus 0.5em minus 0.4em\relax ACM, 2020. [Online].
  Available: \url{https://doi.org/10.1145/3342195.3387517}
\BIBentrySTDinterwordspacing

\bibitem{alibaba-2017}
\BIBentryALTinterwordspacing
Q.~Liu and Z.~Yu, ``{The Elasticity and Plasticity in Semi-Containerized
  Co-Locating Cloud Workload: A View from Alibaba Trace},'' in
  \emph{Proceedings of the ACM Symposium on Cloud Computing}, ser. SoCC
  '18.\hskip 1em plus 0.5em minus 0.4em\relax ACM, 2018, p. 347–360.
  [Online]. Available: \url{https://doi.org/10.1145/3267809.3267830}
\BIBentrySTDinterwordspacing

\bibitem{alibaba-2018}
\BIBentryALTinterwordspacing
H.~Tian, Y.~Zheng, and W.~Wang, ``{Characterizing and Synthesizing Task
  Dependencies of Data-Parallel Jobs in Alibaba Cloud},'' in \emph{Proceedings
  of the ACM Symposium on Cloud Computing}, ser. SoCC '19.\hskip 1em plus 0.5em
  minus 0.4em\relax ACM, 2019, p. 139–151. [Online]. Available:
  \url{https://doi.org/10.1145/3357223.3362710}
\BIBentrySTDinterwordspacing

\bibitem{azure-functions}
Azure, ``{Azure Functions documentation},'' 2021,
  https://docs.microsoft.com/en-us/azure/azure-functions.

\bibitem{azure-containers}
------, ``{Azure Kubernetes Service},'' 2021,
  https://azure.microsoft.com/en-ca/services/kubernetes-service.

\end{thebibliography}

\end{document}